\documentclass[12pt, english]{article}
\usepackage{babel}
\usepackage{epsfig}

\begin{document}

\title
{Search for tribaryon production  in alpha-particles interactions
with protons at intermediate energies}
\author
{A.V.Blinov{\footnote{E-mail: blinov@itep.ru}}, M.V.Chadeyeva{\footnote{E-mail: marina.chadeyeva@itep.ru}}\\
{\it Institute for Theoretical and Experimental Physics}\\
 {\it Moscow, Russia}}
\maketitle

\begin{abstract}
The analysis of the data on the reactions $^4$Hep $\rightarrow$
pppnn and $^4$Hep $\rightarrow$ dppn obtained at the 2-m ITEP
liquid-hydrogen bubble chamber exposed to beams of $^4$He nuclei
with momenta of 2.7 and 5~GeV/$c$ revealed a narrow structures in
the effective-mass spectra of the trinucleon system (NNN) at
2.99~GeV (for isospin $T=3/2$) as well as at 3.04~GeV ($T=1/2$).
The masses of the observed structures are consistent with the
masses of low-lying tribaryon resonances predicted by some
theoretical models. Possible resonance nature of the structures
observed is discussed.
\end{abstract}

The recent discovery in KEK of the narrow strange ($S = -1$)
tribaryons $S^0$(3115) and $S^+$(3140) \cite{Suzuki} i.e.
9q-resonances as predicted by the quark bag models \cite{de
Swart,Maezawa} has given a new impetus to a further search for
other similar candidates,  $^3\!B$ resonances with $S = 0$ in
particular. Previous indications on possible manifestations of the
non-strange tribaryons have been found in some experiments
\cite{Abdinov, Blinov1} but in some others  these states have not
been observed \cite{Shach,Aslanides,Abdullin,Blinov2,Chuvilo}. The
recent quark bag model calculations \cite{Maezawa} predict the
low-lying $^3\!B(S = 0)$ states with the masses as follows:
$M_{{^3}\!B} =$3.1~GeV (for isospin $T = 3/2$) and 3.04~GeV ($T =
1/2$).

As it was first mentioned in ref.\cite{Kondratyuk_p} (see also
refs.\cite{Blinov2,Chuvilo}), the tribaryons predicted by the
simplest generalization of McGregor's rotation model for $^2\!B$
resonances \cite{McGregor} as the rotational excitations of the
NNN$\pi$ system can have the following minimum masses (in GeV):
2.96(L=0), 2.99(L=1), 3.04(L=2), 3.12(L=3),... (L-- is the orbital
angular momentum quantum number of the system). Thus, this simple
model also predicts the existence of the low-lying tribaryons with
masses below the NN$\Delta$(1232) threshold which is $\sim
3.11$~GeV. In this paper we present the results of the very first
attempt to track down such low-lying states by analyzing the mass
spectra of 3p and ppn systems produced in alpha-particles
interactions with protons at intermediate energies.

The results presented are based on the data of the reactions
\begin{equation}
^4{\rm Hep} \rightarrow {\rm pppnn}  \label{eq1}
\end{equation}
and
\begin{equation}
^4{\rm Hep} \rightarrow {\rm dppn}  \label{eq2}
\end{equation}

The isospin of the 3p system from the reaction (\ref{eq1}) is
evidently $T = 3/2$. The reaction (\ref{eq2}) is a unique one
because, as we may see, the ppn system in this case is in the $T =
1/2$ state only. Thus, we have a lucky opportunity to investigate
the trinucleon system in two different pure isotopic states.

The data were obtained at the ITEP 2m liquid-hydrogen bubble
chamber (LHBC) exposed to separated beam of $\alpha$-particles at
2.7~GeV/$c$ and 5~GeV/$c$  (the kinetic energies of primary
protons in the rest system of the nucleus are $T_{\rm p} =
220$~MeV and 620~MeV  respectively). The chamber was placed in a
magnetic field of 0.92~T. The primary-beam background particles
(mainly deuterons) were easily separated from $^4$He nuclei by
visual estimation of the track ionization. About 60,000 pictures
on the 2.7~GeV/$c$ beam and about 120,000 pictures on 5~GeV/$c$
beam were obtained with an average of about 5-8 initial particles
for the chamber extension. About 18,000-19,000 events were
measured at each initial momentum. A more detail description of
the experimental and data processing procedure used in the present
experiment can be found in \cite{Abdullin2, Blinov3}. Note that
the experimental technique applied permits to analyze the data on
reactions (\ref{eq1}) and (\ref{eq2})  in $4\pi$-geometry.

The total $^4$Hep- cross section has the standard form
\cite{Abdullin2} and is determined by the account of the number of
interactions in the fiducial volume. The total cross section is
equal to $109.4 \pm 1.8$ and $121.5 \pm 2.9$~mb at 2.7 and
5~GeV/$c$ respectively (the errors are statistical only). The
systematic error in the absolute normalization of the cross
section is $\sim$3\%.

For particle identification in the case of three-prong events we
used the selection procedure which is standard for bubble chamber
experiments and which takes into account the secondary track
ionization measurements. The events of the reaction (\ref{eq2})
with only one neutral particle in the final state make the
kinematics balance possible. They are easily separated at
5~GeV/$c$ from the events of the channel
$^4$Hep$\rightarrow$dppn$\pi ^0$ (see \cite{Blinov3}). The events
of the reaction (\ref{eq1}) with two neutral particles are
unbalanced. Note that the pion production in $^4$Hep- interaction
is negligible at 2.7~GeV/$c$ ($T_{\rm p} =  220$~MeV) below the
pion production threshold in the elementary NN process. The
channel (\ref{eq1}) at 5~GeV/$c$  has some  evident admixture of
the events of the reaction $^4$Hep$\rightarrow$pppnn$\pi ^0$ which
we estimate as $\sim 5$\%.

The cross sections and the number of events in each channel
(\ref{eq1}) and (\ref{eq2}) for two values of the  initial
momentum are presented in Table 1 (only statistical errors are
indicated).

Note that the use of the $^4$He nucleus in the present experiment
(as well as in the KEK experiment) looks extremely important since
the wave functions of the separate nucleons in this nucleus (the
most compact one) are strongly overlapped and possible multi-quark
effects (valid for small ranges) are most possible.

Fig.1 provides some examples of the diagrams describing the
reactions: $^4{\rm Hep} \rightarrow$ pppnn (quasi-elastic
scattering  (a), quasi-elastic charge exchange reaction (b),
possible production of tribaryon decaying into 3p (c)) and $^4{\rm
Hep} \rightarrow$ dpnn (direct and charge exchange channel (d),
neutron pick-up (e), possible production of tribaryon decaying
into ppn (f)). Here $p_{in}$ is the incident proton, $p_{F},
n_{F}$, and $d_{F}$ are the fast (in the nuclear rest frame)
secondary proton, neutron, and deuteron respectively.

Fig.2 shows the effective mass distribution of 3p system $M_{3p}$
for the reaction (\ref{eq1}) at  2.7~GeV/$c$ (a-c) and 5~GeV/$c$
(d-f). In the distribution at 2.7~GeV/$c$ without any selections
(Fig.2a) there are no statistically confident peaks while at
5~GeV/$c$ there is a narrow structure at $M_{3p}$ $\approx$
2.99~GeV with the statistical significance of $\sim$3.5 standard
deviations from the background (58 events in peak maximum). The
significance is defined as the signal enhancement over the
background divided by its statistical error taking into account
the uncertainty of the background. Evidently, it is a strong
function of the assumed background shape. To suppress the
background contribution of the diagram 1a in reaction (\ref{eq1})
we used the only one selection on the basis of the two-neutron
effective mass $M_{2n} > 1.93$~GeV (for 2.7~GeV/$c$, see Fig.2b)
and $M_{2n} > 2.025$~GeV (for 5~GeV/$c$, Fig.2e). Then the
contribution of the diagram 1b is apparently enhanced.

For comparison, Fig.2c,f presents the distributions with
selections $M_{2n}< 1.93$~GeV (for 2.7~GeV/$c$) and $M_{2n} <
2.025$~GeV (for 5~GeV/$c$). The arrows in Fig.2 show the positions
of the tribaryon masses predicted by the model
\cite{Kondratyuk_p}. But again there are no peculiarities in the
distribution at 2.7~GeV/$c$. As seen in Fig.2b the shape of the
$M_{3p}$ distribution for proton-spectators (in the nucleus rest
frame) is well reproduced by the background distribution received
through a random mix of the events (dotted histogram).

At 5~GeV/$c$ the structure observed in the total spectrum at
$M_{3p}$$\approx$2.99~GeV doesn't change its position with large
2n mass selection (Fig.2e) and is not reproduced by the background
distribution (dotted histogram). Instead of the selection $M_{2n}
> M_{max}$ it is possible to use the selections $M_{2n} > M_{p_ip_k}$ and
$P_{2n} > P_{p_ip_k}$ to suppress the background contribution of
the diagram 1a (here $M_{p_ip_k}$, $P_{p_ip_k}$ are the effective
mass and the total momenta (absolute value in the nucleus rest
frame) of the proton pair $p_ip_k$ (i,k =1,2,3) respectively).
The number of events selected using the restrictions mentioned
above is practically the same: 394 (for $M_{max} = 2.025$~GeV),
403 (for fast $M_{2n}$ selection), and 404 (for fast $P_{2n}$
selection). The structure at $M_{3p}$ $\approx$ 2.99~GeV appears
under these selections as well.

The position of the maximum revealed in the present experiment is
in close agreement with the model predictions \cite{Kondratyuk_p}
for the P-wave ($L=1$) tribaryon (it is evident that the S-wave 3p
state at rest is forbidden by the Pauli exclusion principle). But
to make sure of  the resonance nature for this structure (see
diagram 1c), however, it would be important to have a higher
statistics and also to understand better the contribution of the
mechanisms with the production of pions in the intermediate state
followed by their capture by the correlated NN pair (note that the
position of the structure is around the mass of $NNN\pi$). And
yet, the partial wave analysis of the angular distributions is
needed to determine the quantum numbers of the assumed tribaryon.

The data fit in the interval $2.86$~GeV $\le M_{3p} \le 3.09$~GeV
by the sum of Breit-Wigner function and the polynomial background
gives the following parameters (mass and width) of the given
structure: $M_{X}= 2.99 \pm 0.01$~GeV and $\Gamma_{X} = 0.024 \pm
0.017$~GeV ($\chi^2/\rm{NDF} = 0.14$). The solid line in Fig.2d
shows the result of the fit (the background is marked by the
dash-dotted line). The production cross section of the possible
coupled state is $\approx 0.43 \pm 0.22$~mb (only statistical
error is indicated). The solid line in Fig.2e shows the fit of the
data in the interval $2.89$~GeV $\le M_{3p} \le 3.1$~GeV by the
sum of Breit-Wigner function with the parameters $M_{X} = 2.99 \pm
0.01$~GeV , $\Gamma_{X} = 0.04 \pm 0.02$~GeV and the exponential
approximation (dashed curve) for phase space ($\chi ^2 / \rm{NDF}
= 0.43$). These fit parameters are in agreement with the ones for
the total spectrum within the statistical uncertainties. A random
mix of events is shown in Fig.2e by the dotted histogram. In the
mass region of interest this background is close to the
exponential one (see Fig.2e).

To determine the own (natural) mass and width of the assumed
tribaryon  we use the modification of the usual Breit-Wigner
function with due regard to experimental resolution (see, i.e.
\cite{Blinov3}). The mean absolute resolution for $M_{3p}$ at
5~GeV/$c$ is $\sigma$ $[M_{3p}]$ $\le 19 $MeV in the region of
interest. It is compared to the fitted value of $\Gamma_{X}$. The
fit of the data with the account of the experimental resolution
gives almost the same  estimation for the width of the assumed
tribaryon in the case of total $M_{3p}$ distribution  as well as
for the spectrum under restriction: $\Gamma_{{^3}\!B}$ $\le$ 66
MeV (95\% C.L.). Note that the maximum position doesn't depend on
the uncertainties of the background and its approximation.

Fig.3 shows the effective mass distribution of ppn system
$M_{ppn}$ for the reaction (\ref{eq2}) at  2.7~GeV/$c$ (a-c) and
5~GeV/$c$ (d-f). The arrows in Fig.3 show the positions of the
tribaryon masses predicted by the model \cite{Kondratyuk_p}.  In
the distributions  without any selections for both initial momenta
there are no statistically confident  peaks (see Fig.3a,d).

To suppress the background contribution of the  diagram 1d in the
reaction (\ref{eq2}) we used the selections on the deuteron
momentum in the $^4$He rest frame $P_{d} > 0.6$~GeV (for 2.7
~GeV/$c$, Fig.3b) and $P_{d} > 1.1$~GeV (for 5~GeV/$c$, Fig.3e).
Then the contribution of the diagram 1e will be apparently
enhanced. For comparison, the distribution with selections $P_{d}<
0.6$~GeV (for 2.7~GeV/$c$) and $P_{d}< 1.1$~GeV(for 5~GeV/$c$) are
presented in Fig.3c and Fig.3f respectively. There are no
peculiarities in spectra at 2.7~GeV/$c$. At 5~GeV/$c$ there is a
narrow structure at $M_{ppn} \approx 3.04$~GeV with the
statistical significance of $\sim$3.5 standard deviations from the
background (20 events in the peak maximum) which may be the
indication of the possible $^3\!B$ resonance production in the ppn
system (diagram 1f). The cut value choice doesn't affect the
distribution shape at 2.7~GeV/$c$. At 5~GeV/$c$, the cut value has
been chosen to maximize the effect while having enough statistics.

The fit of the data in the interval $2.9$~GeV $\le M_{ppn} \le
3.09$~GeV by the sum of Breit-Wigner function and the exponential
background gives  the mass and  the width of the observed
structure $M_{Y}= 3.039 \pm 0.005$~GeV and $\Gamma_{Y} = 0.035 \pm
0.008$~GeV ($\chi^2/\rm{NDF}=0.18$). The solid line in Fig.3e
shows the result of the fit (the background is marked by the
dash-dotted line). Note, the fit of the data in the same interval
by the exponential function only leads to the value
$\chi^2/\rm{NDF}=2.1$, which does not permit to treat this
approximation as statistically significant. The position of the
maximum is in close agreement with the simple model predictions
\cite{Kondratyuk_p} for the D-wave ($L=2$) tribaryon as well as
with the quark bag model calculations \cite{Maezawa} for $T=1/2$
tribaryon. The production cross section of the possible coupled
state is $\approx 0.25 \pm 0.08$~mb (only statistical error is
indicated). The mean absolute resolution for $M_{ppn}$ at
5~GeV/$c$ in the region of interest is $\sigma$ $[M_{ppn}]$ $\le
31$ MeV  which gives  the estimated width of the assumed
tribaryon: $\Gamma_{^3\!B^*}$ $\le$ 60~MeV (95\% C.L.).

The main results of the paper are as follows.

The narrow structures were seen in the trinucleon mass spectra in
the reactions $^4{\rm Hep} \rightarrow$ pppnn and  $^4{\rm Hep}
\rightarrow$ dppn. The positions of the observed structures are in
a very good agreement with the simple rotation model predictions
for low-lying non-strange tribaryons where they are interpreted as
the rotation excitations of the NNN$\pi$ system. Furthermore, the
position of the structure for $T=1/2$ is also consistent with 9q
quark bag model calculations. But to make sure of  the resonance
nature for these structures it is necessary to understand better
the contribution of the mechanisms with the production of pions
and deltas in the intermediate state.

It looks rewarding in the future to investigate the trinucleon
mass spectra in the reactions (\ref{eq1}) and (\ref{eq2}) in the
same mass region but at greater initial energies and with a higher
statistics. It might appear  promising as well  to study the
possible decay mode into the NNN$\pi$ in the mass region of
(3N$\pi$, NN$\Delta$).

The authors are pleased to thank V.F.~Turov for his assistance in
data processing procedure, V.V.~Kulikov for fruitful discussions,
and A.G.~Dolgolenko for correct critical comments and constructive
proposals.

This work is supported by the grant no. 04-02-16500 of the Russian
Foundation for Basic Research.

\begin{figure}[p]
\vskip 5cm \epsfysize=10cm \centerline{\epsffile{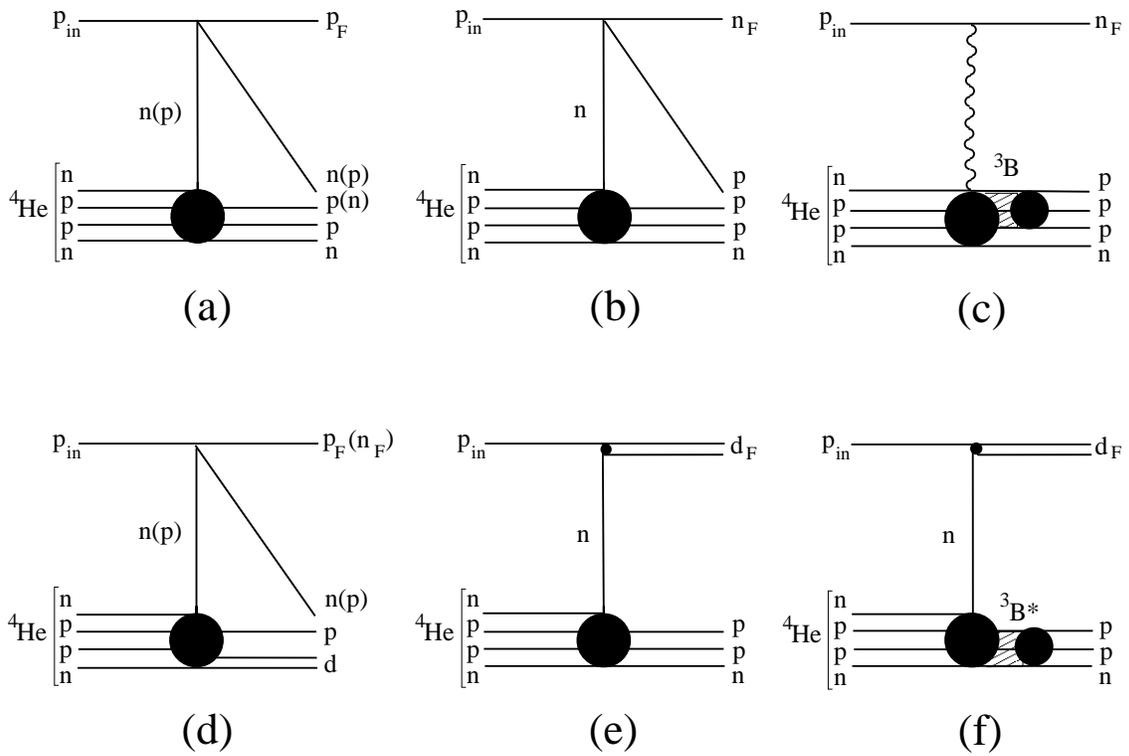}}
\caption{\small Some of possible diagrams describing the
reactions: p$^4$He $\rightarrow$ pppnn (quasi-elastic scattering
(a), quasi-elastic charge exchange reaction (b), possible
production of tribaryon decaying into 3p (c)) and p$^4$He
$\rightarrow$ dppn (direct and charge exchange channel (d),
neutron pick-up (e), possible production of tribaryon decaying
into ppn (f)).} \label{fig1}
\end{figure}

\begin{figure}[p]
\vskip -2cm \epsfysize=14cm \centerline{\epsffile{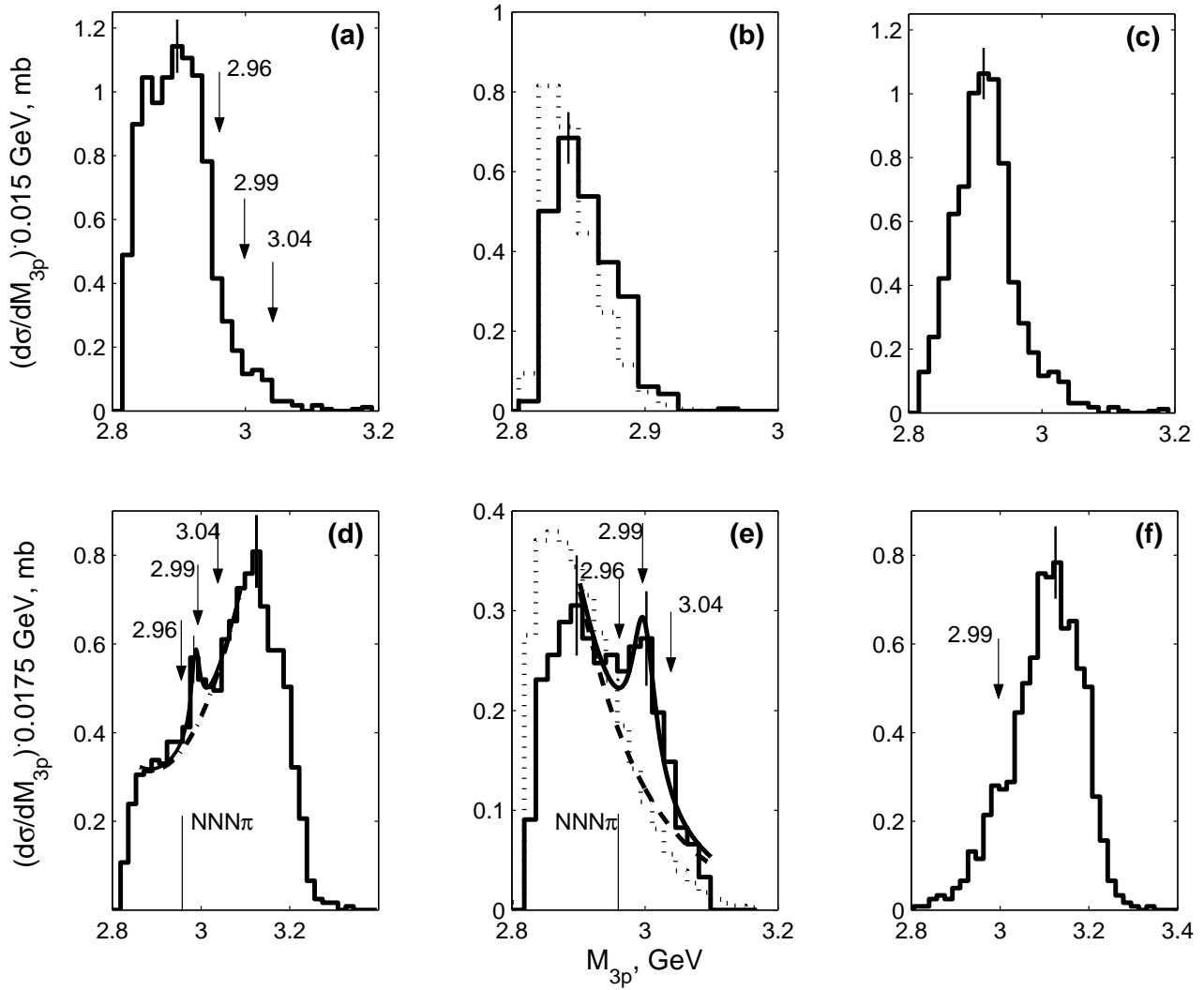}}
\caption{\small The effective mass distribution of 3p system
$M_{3p}$ for the reaction $^4$Hep $\rightarrow$ pppnn at $^4$He
momentum of 2.7~GeV/$c$ (all events - (a),  the events selected by
two-neutron effective mass: $M_{2n} > 1.93$~GeV - (b), $M_{2n} <
1.93$~GeV - (c)) and 5~GeV/$c$ (all events - (d),  the events
selected by two-neutron effective mass: $M_{2n} > 2.025$~GeV -
(e), $M_{2n} < 2.025$~GeV- (f)). Solid curves represent the fit
for experimental data as a sum of the background (dash-dotted
curve in figure (d) and dashed -- on figure (e)) and Breit-Wigner
function (see text). Dotted histograms correspond to the
experimental background distribution received through a random mix
of the events.} \label{fig2}
\end{figure}

\begin{figure}[p]
\vskip -2cm \epsfysize=14cm \centerline{\epsffile{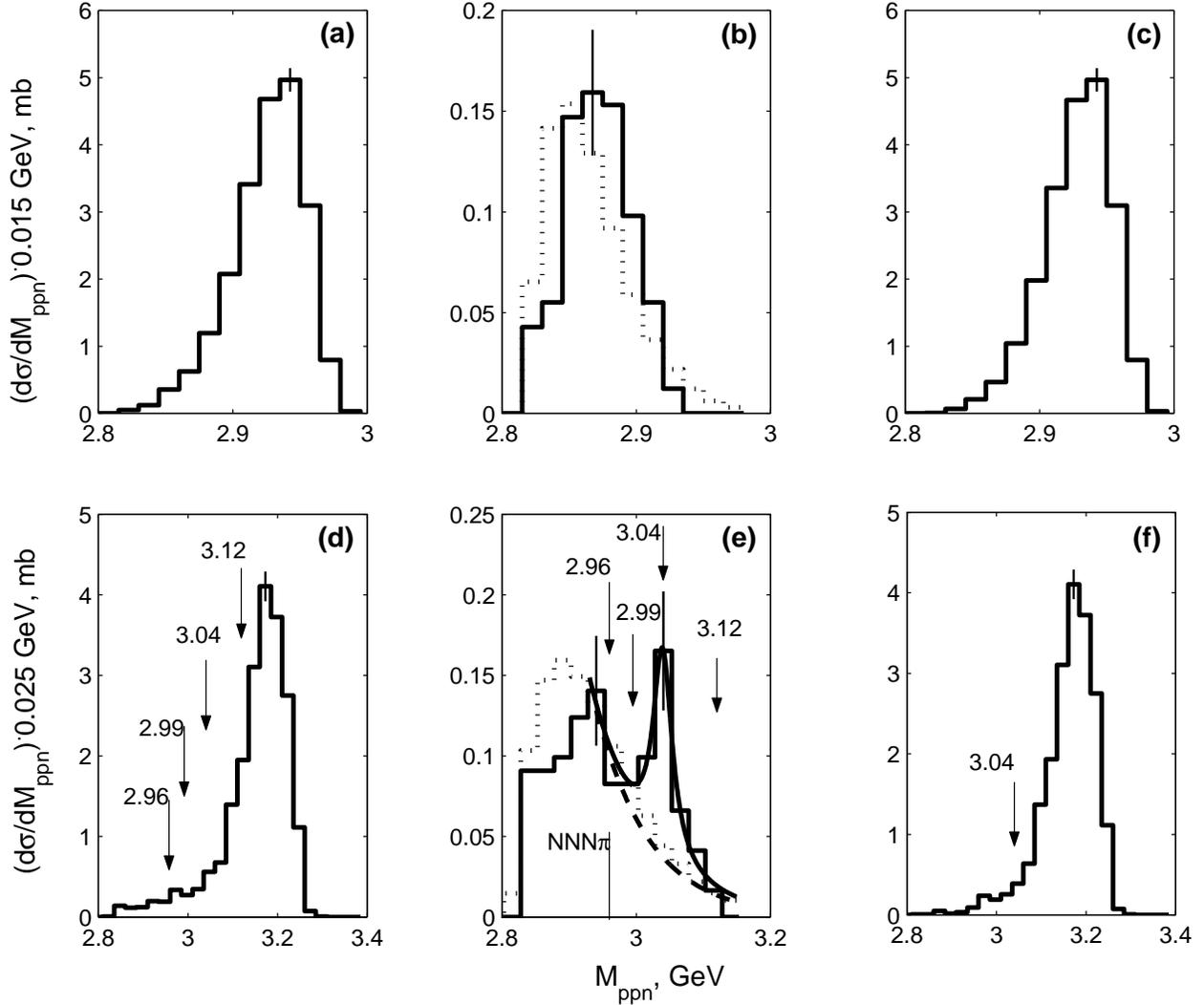}}
\caption{\small The effective mass distribution of ppn system
$M_{ppn}$ for the reaction $^4$Hep $\rightarrow$ dppn at $^4$He
momentum of 2.7~GeV/$c$ (all events - (a), the events selected by
deuteron momentum  in the nucleus rest frame: $P_{d} >
0.6$~GeV/$c$ - (b), $P_{d} < 0.6$~GeV/$c$ - (c)) and 5~GeV/$c$
(all events - (d), the events selected by deuteron momentum in the
nucleus rest frame: $P_{d} > 1.1$~GeV/$c$ - (e), $P_{d} <
1.1$~GeV/$c$ - (f)). Solid curve represents the fit for
experimental data as a sum of the background (dashed curve) and
Breit-Wigner function (see text). Dotted histograms correspond to
the experimental background distribution received through a random
mix of the events.} \label{fig3}
\end{figure}

\begin{table}[tbh]
\caption {\small Cross sections of the reactions (\ref{eq1}) and (\ref{eq2}) at
$2.7$~GeV/$c$ ($T_{\rm p} = 220$~MeV) and $5$~GeV/$c$ ($T_{\rm p}
= 620$~MeV) initial momenta.}
\begin{center}
\begin{tabular}{|c|c|c|r@{$\pm$}l|}
\hline
 Momentum, &  Channel & Number of events & \multicolumn{2}{c|}{Cross section,} \\
 GeV/$c$ & & & \multicolumn{2}{c|}{mb} \\ \hline
2.7 & $^4$Hep$\rightarrow$dppn  & 3474 &  21.4 & 0.4 \\ \cline{2-5}
  & $^4$Hep$\rightarrow$pppnn & 1620  & 9.9 & 0.2 \\ [5 pt] \hline
 5 & $^4$Hep$\rightarrow$dppn & 2567 & 21.2 & 0.4 \\ \cline{2-5}
  & $^4$Hep$\rightarrow$pppnn & 1394  & 11.5 & 0.3 \\ [5 pt] \hline
\end{tabular}
\end{center}
\end{table}
\end{document}